\documentclass[%
 reprint,
 amsmath,amssymb,
 aps,
]{revtex4-2}

\usepackage{graphicx}
\usepackage{dcolumn}
\usepackage{bm}
\usepackage[inkscapelatex=false]{svg}
\usepackage{hyperref}

\usepackage{todonotes}
\usepackage{changes}
\usepackage{xcolor} 
\setuptodonotes{inline}
\begin{document}

\title[Extraction of Effective Electromagnetic Material Properties for Rydberg Electrometer Vapor Cells from 10-300 MHz]{Extraction of Effective Electromagnetic Material Properties for Rydberg Electrometer Vapor Cells from 10-300~MHz\\}
\author{D. Richardson}
\author{J. Dee}
\author{J. Yaeger}
\author{M. Viray}
\author{J. Marsh}
\author{B. Kayim}
\author{B. C. Sawyer}
\author{D. S. La Mantia}
\author{R. Wyllie}
\author{R. S. Westafer}
\affiliation{ 
Georgia Tech Research Institute, USA 
}%


\date{\today}

\begin{abstract}
Quantum sensors often consist of packaging, such as dielectric-based vapor cells and metallic electrodes, that reduces and spatially alters the locally observed electromagnetic fields. These effects have been well studied in the optical regime, and even in the RF regime over a few~GHz. However, there have been few studies in the electrically small regime below 1~GHz. In order to account for or remove the effects of the packaging, more studies are needed across a broad range of frequencies. This paper reports on the complex permittivity and conductivity of several commercially available vapor cells used for Rydberg electric field sensing from 10-300~MHz. A new method using a stripline transmission measurement was performed and full wave electromagnetics modeling was used to extract the effective dielectric constitutive parameters from the vapor cells. Additionally, the field reduction inside the vapor cell is reported, and published atomic measurements of the electric field are used to further validate the results presented here. Several observations were made from the measurements, such as the frequency dependencies of the RF dispersion and absorption. Applications of this technique include making precise numerical field corrections or physically designing a more optimal vapor cell via coatings, material changes, or geometric changes to improve field strength and uniformity.
\end{abstract}

\maketitle

\section{\label{intro}Introduction\protect\\}
Rydberg atom receivers or field sensors typically consist of atomic vapor, normally rubidium or cesium, placed inside a hermetically-sealed dielectric enclosure, such as glass or Si-glass composites, and are commonly referred to as vapor cells \cite{cells1,cells2}. Several schemes have been adopted to raise the outermost electrons of these atomic vapors into the high energy Rydberg states that are sensitive to incident RF fields. Most consist of two or more all-optical beams, and some rely on local RF oscillators (LO) in addition to optical beams \cite{shaffer_resub,shaffer_resub2,PhysA_LindbladEq,heterodyne,review_paper}. These quantum sensors have the potential to supplement or replace traditional RF sensing technologies due to the high sensitivity, electrically small packaging, and unconventional field detection mechanisms they can provide \cite{highsensitivity,other_uses,usefulness}. When compared with traditional RF sensing technologies, these devices interact differently with incident fields, acting as active receivers and RF mixers \cite{mixer}. These sensors have the potential to measure the electromagnetic fields with unrivaled accuracy and precision when compared with traditional field sensors, due to the fact that they are very responsive to local RF fields while minimally perturbing said local fields \cite{single_photon}. However, the packaging for these quantum sensors (and some atomic effects) reduces and spatially alters the fields. In the non-interacting scenario, the atomic vapor would be close to the permittivity of vacuum, while the glass material would be that of a low dielectric material \cite{cell_interaction}. However, atomic vapor interacting with the vapor cell walls creates a stronger shielding effect \cite{sandia_new,sandia_old}. The shielding (as a function of frequency) leads to reduced field strength and uniformity degradation that will cause a reduction in sensitivity, as well as substantial angle of arrival estimation errors when using an array of Rydberg field sensors for direction finding \cite{QA_array,bob_paper}. The properties of vapor cells have been well studied in the optical regime to allow high transmission of optical laser excitation in vapor cells, and there have been some studies in the kHz regime up to the low MHz regime \cite{sandia_new,sandia_old,optical_sh1,optical_sh2}. However, there have been no studies (to the knowledge of the authors) in the electrically small regime from 10~MHz up to 1~GHz. Given the importance of improving quantum sensing devices and the lack of experimental studies to quantify the RF material properties of broadband Rydberg sensors, more studies are needed across a range of frequencies for a wide variety of vapor cell designs in order to account for or remove the effects of the packaging.

This paper presents an experimental and numerical effort to quantify the effective complex RF material properties of several commercial-off-the-shelf (COTS) vapor cells from 10-300 MHz. A new method is used that utilizes a stripline transmission line transmission measurement with full wave electromagnetics modeling to determine the effective complex RF material properties. Several vapor cells are considered for this study: an unfilled quartz glass vapor cell, a rubidium87-filled quartz glass vapor cell, a rubidium-filled quartz glass vapor cell, a rubidium-filled sapphire vapor cell, an unfilled borosilicate glass vapor cell, a rubidum-filled borosilicate glass vapor cell, two sodium-filled borosilicate glass vapor cells, and two cesium-filled borosilicate glass vapor cells. The rubidum87-filled quartz vapor cell is of high purity ($>98\%$), while the other rubidium cell contain an abundance of other isotopes. Unless otherwise stated, the material property extraction is performed at room temperature ($22~^\circ C$) with alkali-metal in the ground state (no excited Rydberg states). It is challenging from both a model and measurement perspective to treat the vapor cell walls and atomic vapor separately; however, the predominant effect is likely at the vapor cell wall interface. In this paper, the complex RF permittivity and conductivity of the combined materials are quantified as if it were a single, effective material. This approximation is likely valid in the frequency regime studied here. The effective complex RF material properties are quantified for each vapor cell and their respective shielding amounts determined within the context of the stripline transmission line waveguide. In addition to this, several observations are made by comparing the effective complex RF material properties of the vapor cells. Previous studies have concluded the effect described here is solely due to the electrical conductivity of the vapor, but this work suggests conductivity alone can't completely account for or explain the phenomenon. Applications of this technique include making precise numerical field corrections or physically designing a more optimal vapor cell via coatings, material changes, or geometric changes.

\section{\label{methods}Methods\protect\\}
\begin{figure}
\includegraphics[width=\linewidth]{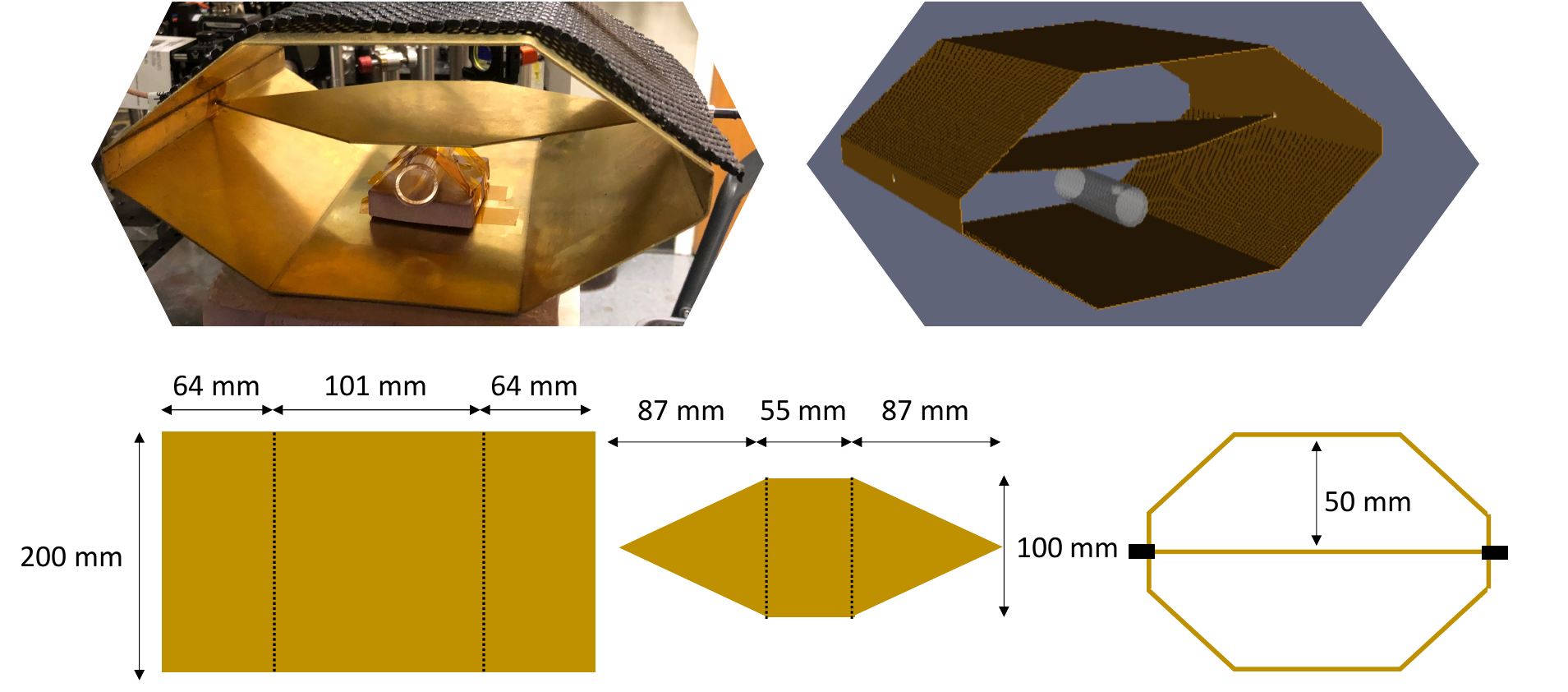}
\caption{Stripline waveguide designed for measuring atomic vapor cells from 10-300~MHz. Both a photo of a prototype and a rendering of the FDTD model are shown.}
\label{fig_stripline}
\end{figure} 
\begin{figure}
\includegraphics[width=\linewidth]{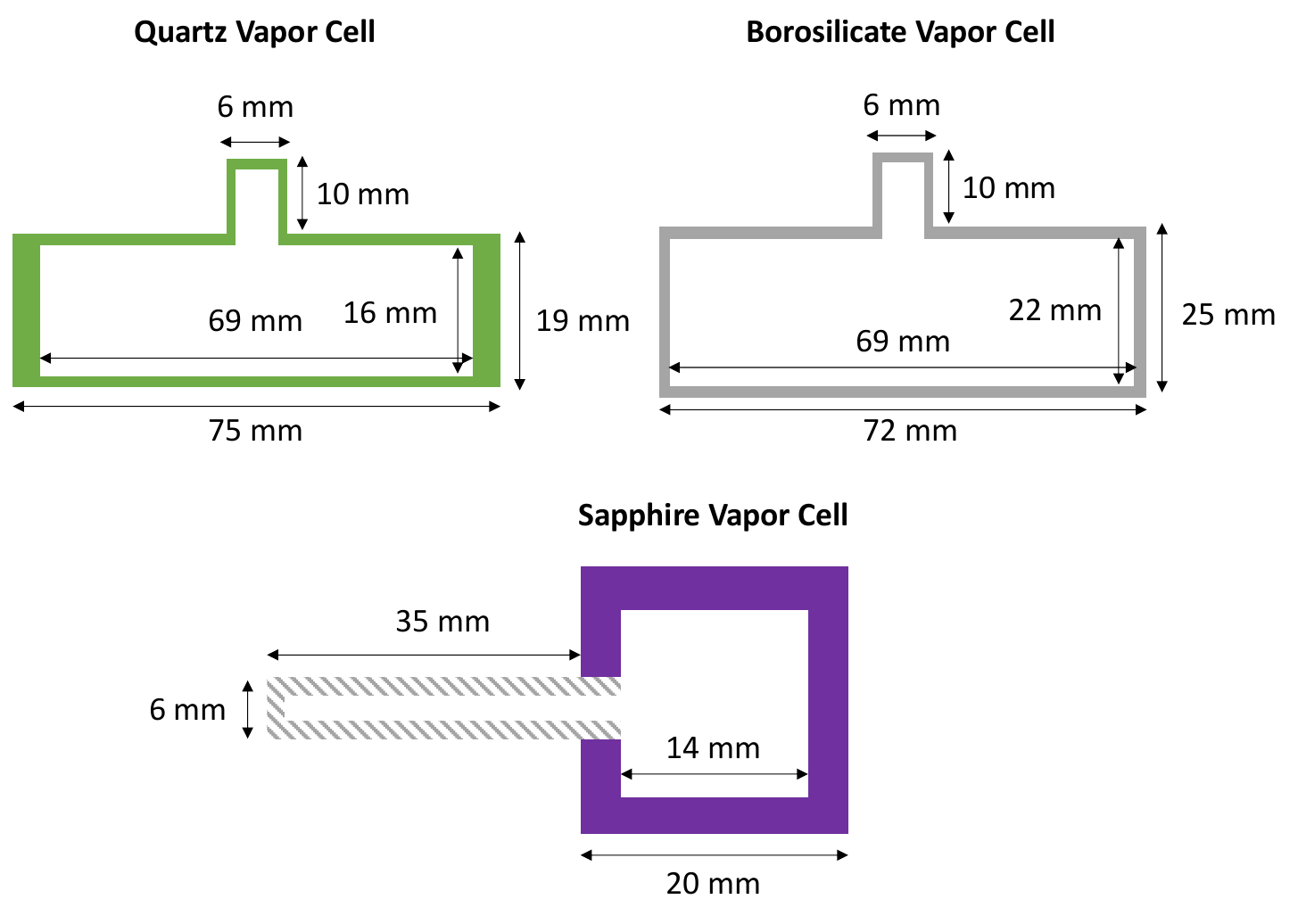}
\caption{Approximate physical dimensions of the various vapor cells. The quartz and borosilicate vapor cells were obtained from Thorlabs, Inc., while the sapphire vapor cell was obtained from Japan Cell Co., Ltd.. The quartz vapor cell is shown in green, the borosilicate vapor cell is shown in grey, and the sapphire vapor cell is shown in purple with a grey borosilicate-based stem.}
\label{fig_cells}
\end{figure}
\begin{figure}
\includegraphics[width=\linewidth]{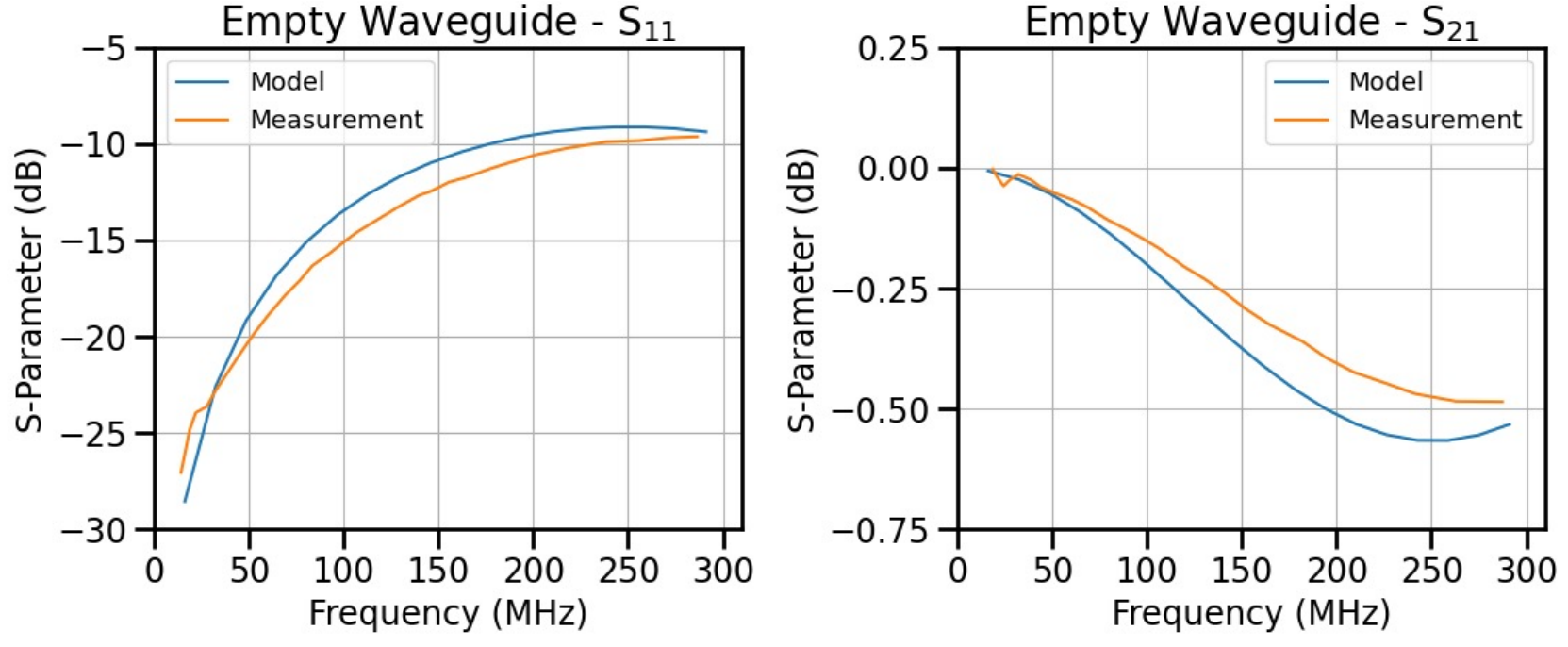}
\caption{Verification of the electromagnetics model by comparison to measurements of the two-port network scattering parameters for the case of the empty waveguide.}
\label{fig_measure_model_empty}
\end{figure}
A stripline waveguide operating in a TEM mode was used for effective complex RF material property extraction. The actual stripline waveguide is shown in Fig.~\ref{fig_stripline}, along with a full wave finite-difference time-domain (FDTD) electromagnetics model \cite{yee,sullivan,taflove}. The critical dimensional features of the waveguide are also displayed in the figure. All of the vapor cells measured here are COTS, as previously stated, and the respective vendors can provide the physical dimensions upon request. The quartz and borosilicate vapor cells were obtained from Thorlabs, Inc., while the sapphire vapor cell was obtained from Japan Cell Co., Ltd.. A simplified drawing of the vapor cells is given in Fig.~\ref{fig_cells}. The quartz vapor cells consist of a hollow cylindrical body with a small stem centered on the cylinder. The overall outer dimensions of the main body are approximately 75~mm in length by 19~mm in diameter, with a wall thickness of 1.5~mm for the main body and close to 3~mm for the end caps.
\begin{figure*}
\includegraphics[width=\linewidth, height=18cm]{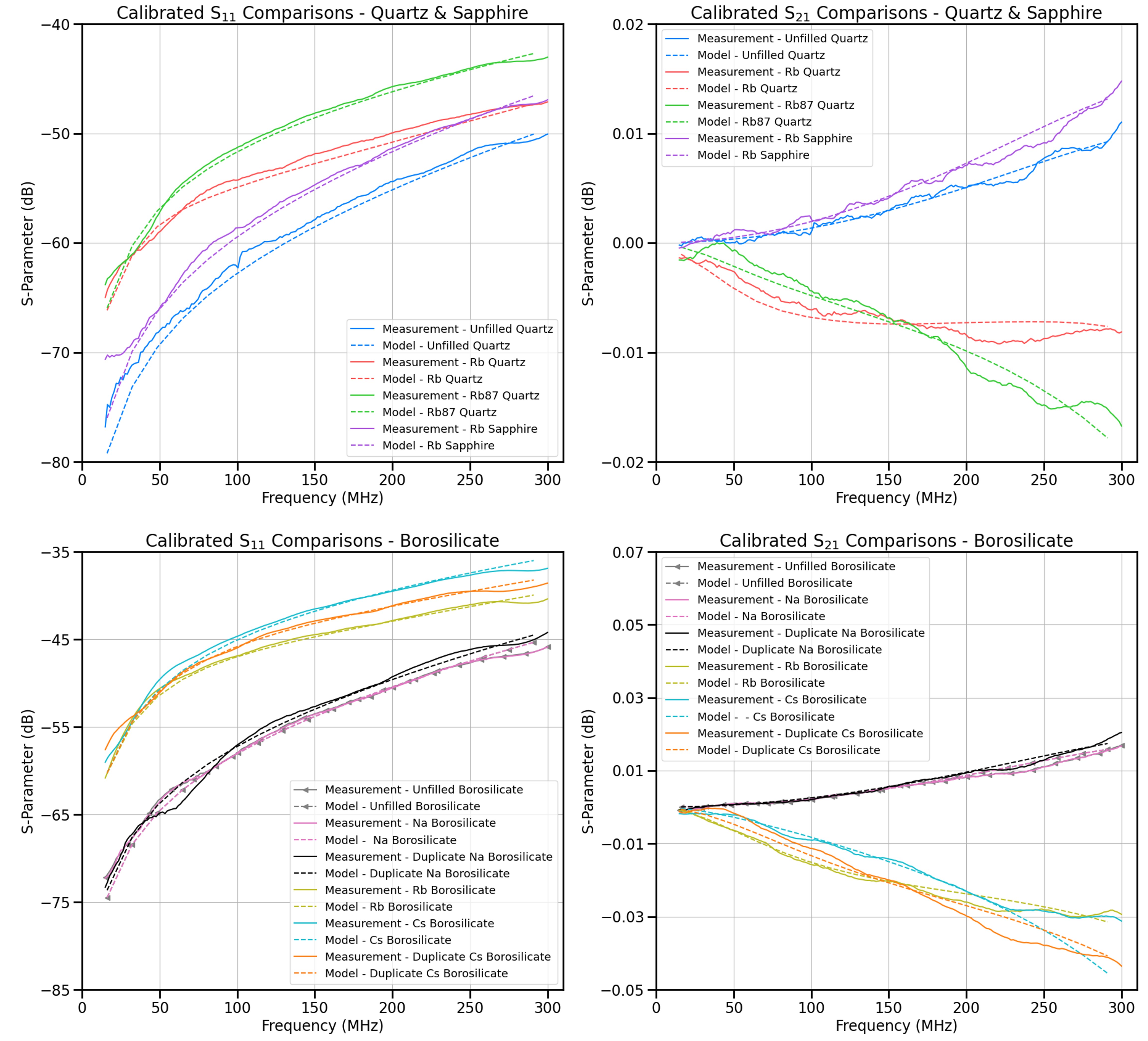}
\caption{Comparison of measured and modeled calibrated scattering parameters $S_{11}$ (left) and $S_{21}$ (right) for several different COTS vapor cells. The plots on the top row show the quartz and sapphire vapor cells, while the plots on the bottom row show the borosilicate vapor cells - which are all separated for ease of viewing.}
\label{fig_S_param_compare}
\end{figure*}
\begin{figure*}
\includegraphics[width=\linewidth, height=6cm]{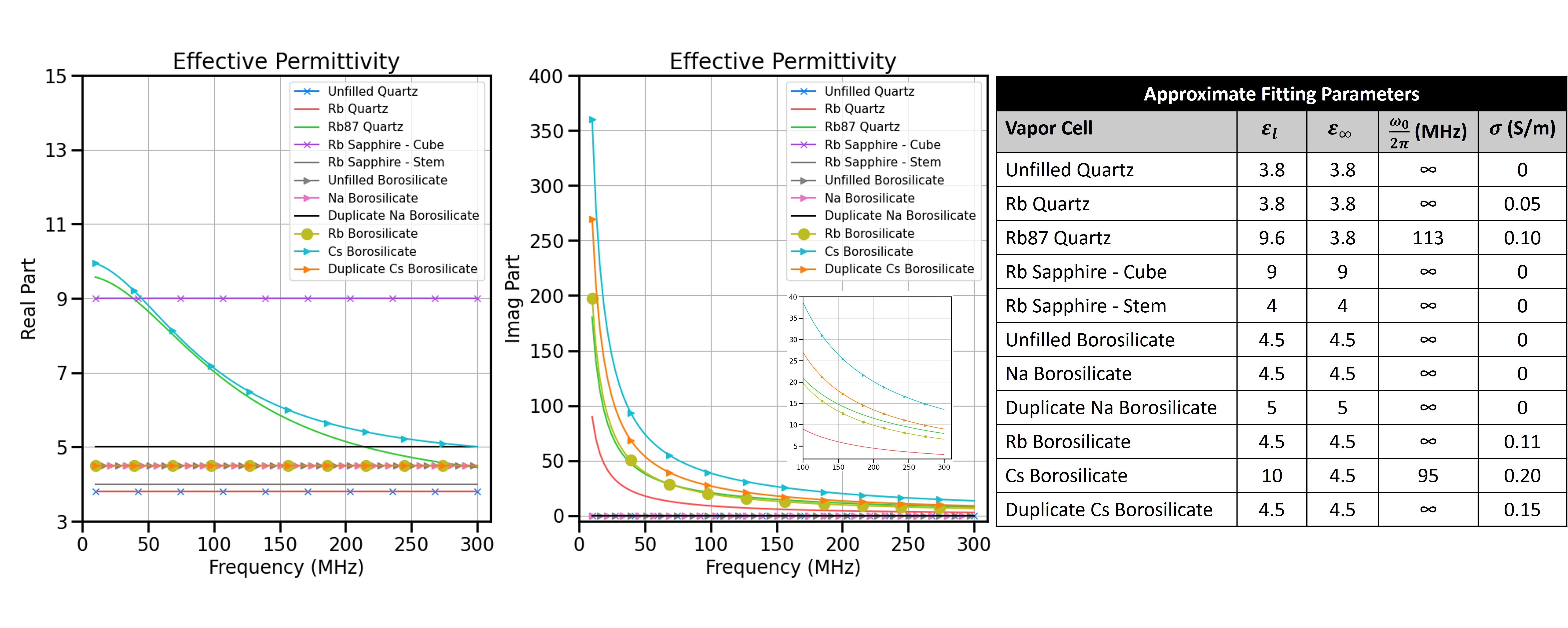}
\caption{Effective complex RF permittivity of the different vapor cells and a table of the fitting parameters used for each vapor cell. Some of the vapor cells didn't require Debye relaxation to obtain a good fit and are better represented by constant dielectric responses, with or without conductive loss. This is represented in the table by an infinite relaxation frequency, equivalent to a relaxation time of zero, with $\epsilon_{\infty}$ = $\epsilon_{l}$.}
\label{fig_Ep_real_and_imag}
\end{figure*}
The borosilicate vapor cells also consist of a hollow cylindrical body with a small stem centered on the cylinder. The overall outer dimensions of the main body are approximately 72~mm in length by 25~mm in diameter, with a wall thickness of 1.5~mm for the main body and close to 1.5~mm for the end caps. The sapphire vapor cell contains a hollow cubic body made from sapphire with a large borosilicate-based stem centered on the cube. The main body outer dimensions are 20x20x20~mm$^3$ and the wall thickness is 3~mm. It should be noted that the quartz vapor cell data sheet provided by Thorlabs, Inc. had some discrepancies and did not completely match the measurements conducted in the lab. Additionally, the borosilicate wall thickness provided by Thorlabs, Inc. was given as an approximate value so it was also measured in the lab. All values measured in the lab are reflected in Fig.~\ref{fig_cells}. The model and measurement S-parameters for the empty waveguide (no vapor cell) are shown in Fig.~\ref{fig_measure_model_empty}, indicating good agreement between the modeled waveguide and the real waveguide. Though not required for material property extraction, the waveguide is highly transmissive and the field largely uniform over the vapor cell in this frequency range. Fig.~\ref{fig_stripline} shows the vapor cell mounted atop a slab of pink foam for support. Measurements were performed with and without the foam, with no noticeable changes in the measured data.

The effective material property extraction procedure contains several important steps. First, calibrated S-parameters are obtained from the model and measurement systems. The calibration effectively removes the waveguide response from the S-parameters. This implies the accuracy in the model and measurement agreement is largely dependent on the accuracy of the cross sectional area of the waveguide dimensions between the two cases. This calibration consists of measuring (and modeling) the empty waveguide, the waveguide with the vapor cell in question, and finally the empty waveguide with an electromagnetic short spanning the entire cross section of the waveguide. Sometimes this is referred to as a TRL (thru, reflect, line) calibration \cite{pozar}. The calibrated S-parameters can be determined by:
\begin{equation}
S_{11}^{cal}=-\frac{S_{11}^{cell}-S_{11}^{empty}}{S_{11}^{short}-S_{11}^{empty}}
\label{eq:cal_S11}
\end{equation}
\begin{equation}
S_{21}^{cal}=-\frac{S_{21}^{cell}-S_{21}^{short}}{S_{21}^{empty}-S_{21}^{short}}
\label{eq:cal_S21}
\end{equation}
Though included in the calibration, the shorted case contributed inconsequential amounts of correction to the calibration because the waveguide transmission is high. Additionally, the S-parameters are time gated to remove additional, unwanted effects from the surrounding environment, as well as some noise. Next, the model and measurement S-parameters are compared to one another. The effective complex RF permittivity and the conductivity of the vapor cell are repeatedly modified in the full wave FDTD model until a good agreement between the model and measurement S-parameter curves is obtained. Once a good agreement is obtained, there is high confidence in the accuracy of the effective complex RF material properties. As an aside, the permittivity (bound charges) and conductivity (free charges) are often reported together as one expression and labeled permittivity only, as will be done in this work. The effective complex RF permittivity can take several forms depending on the causal models used to yield a best fit \cite{jackson}. Using a causal model fit across the broadband gives higher confidence in the extraction. In this work, the properties are allowed to take on the form:
\begin{equation}
\epsilon(\omega) = \epsilon_{\infty} + \frac{\epsilon_l - \epsilon_{\infty}}{1 + j\frac{\omega}{\omega_0}} - \frac{j\sigma}{\epsilon_0\omega}
\label{eq:perm}
\end{equation}
This term includes a Debye model relaxation term and a conductivity term. Additional causal terms can be used if the underlying physics permits, but it was found that this model is mostly sufficient to capture the physics of interest here. $\epsilon_{\infty}$ and $\epsilon_{l}$ represent the permittivity in the high and low frequency limit, respectively. $\omega$ is the angular frequency and $\omega_0$ is the dipole relaxation angular rate ($2\pi{}f_0$ or $2\pi/\tau_0$). Lastly, $\sigma$ is the electrical conductivity.

Once there is good agreement between the model and measurement calibrated S-parameters, the fields inside can be examined and compared to the case of the empty waveguide to quantify the RF electromagnetic shielding. Kayim \emph{et al}. directly used the rubidium87-filled quartz vapor cell and performed atomic spectroscopy to determine the fields observed by the atoms \cite{Baran_paper}. This is briefly discussed below as further evidence of accuracy when using this technique, as well as providing a direct use case.
 \begin{figure*}
\includegraphics[width=\linewidth,height=8cm]{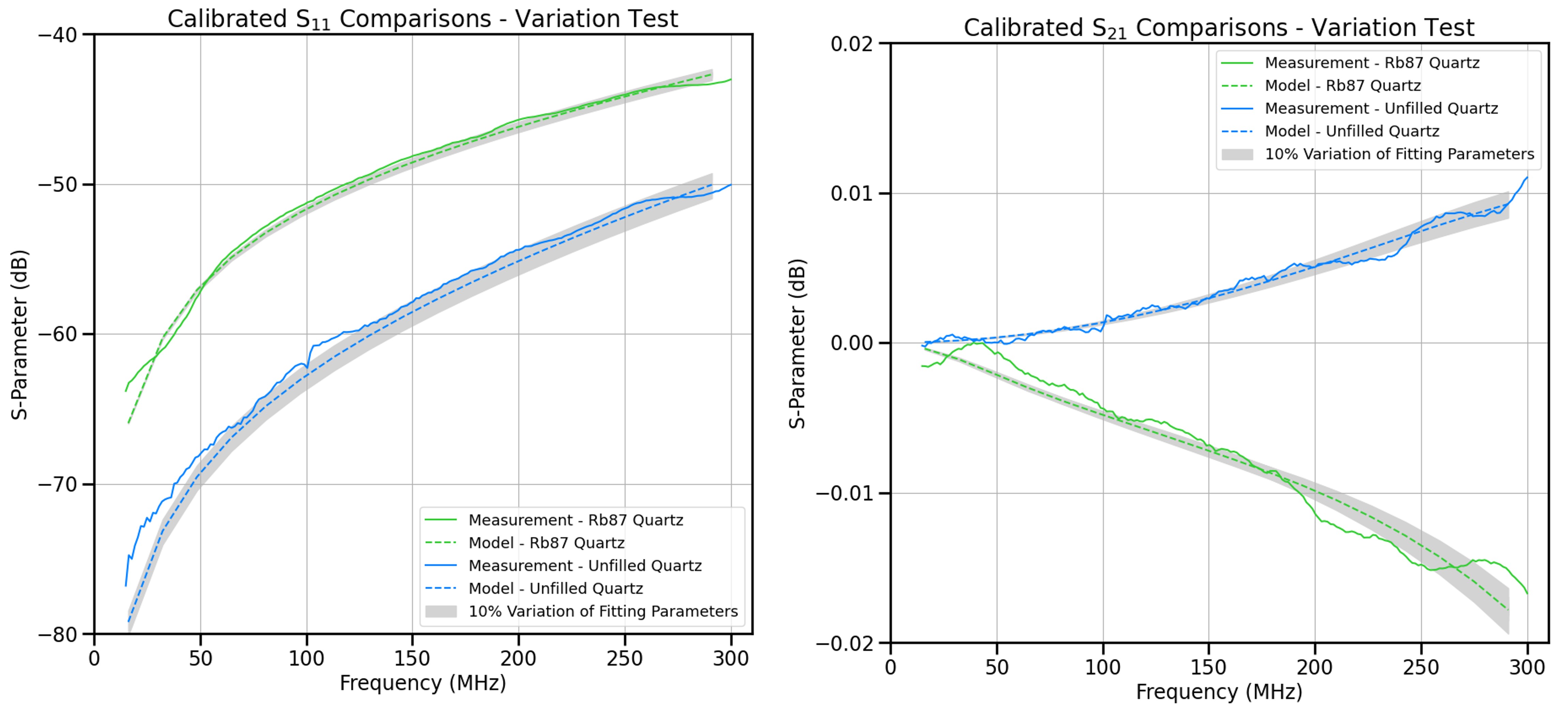}
\caption{Comparison of measured and modeled calibrated scattering parameters $S_{11}$ (left) and $S_{21}$ (right)}. The area produced from a 10$\%$ variation in each of the fitting parameters individually, and combinations thereof, is shown in gray. This indicates the likely accuracy of the fitting parameters used is within 10$\%$, and thus the permittivity as well given it's functional form.
\label{fig_variations}
\end{figure*}

\section{\label{results}Results\protect\\}
\subsection{\label{S-parameter Measurement and Permittivity Determination}S-parameter Measurement and Permittivity Determination}
 The results for the calibrated S-parameters using this new method are shown in Fig.~\ref{fig_S_param_compare}, and the accompanying effective complex RF material property values shown in Fig.~\ref{fig_Ep_real_and_imag}. It should be noted that greater than 0~dB in the calibrated $S_{21}$ does not indicate greater than 100\% transmission. When calibrated $S_{21}$ is greater than 0~dB, it indicates more energy is coupled through the waveguide when the vapor cell is present vs. when it is empty. Interestingly, it is the lower loss and more dielectric-like response vapor cells that have a calibrated $S_{21}$ greater than 0~dB. The unfilled quartz vapor cell showed an effective complex permittivity of 3.8+0j, which is consistent with what would be expected from the bulk properties of quartz glass. The rubidium-filled quartz vapor cell response is dominated by the electrical conductivity produced by the atomic vapor and vapor cell wall interaction over this frequency range. However, the rubidium87-filled quartz vapor cell showed a small dispersive response as well. The measurement of the sapphire vapor cell showed little to no additional contribution from the vapor and vapor cell wall interaction. The measurement yielded an effective complex permittivity of 9+0j for the sapphire portion of the sapphire vapor cell and 4+0j for the borosilicate portion of the vapor cell, consistent with expected bulk dielectric values. Note that only one orientation of the sapphire axis is measured, and while the axis is unknown, it is likely perpendicular to the major c-axis given the lower dielectric constant. The unfilled borosilicate vapor cell, made by Thorlabs inc., showed a slightly larger effective complex permittivity of 4.5+0j. It is not unusual to see variations in glass permittivity from manufacturer to manufacturer. The sodium-filled borosilicate vapor cells both showed responses almost identical to the unfilled borosilicate vapor cell, indicating little to no vapor and vapor cell wall interaction. One of the sodium-filled vapor cells shows a slightly stronger dielectric response, likely due to variations in the physical cell dimensions or variations in borosilicate dielectric constant, both based on manufacturing tolerances. The rubidium-filled borosilicate vapor cell and one of the cesium-filled borosilicate vapor cells are dominated by a conductive response, while one cesium-filled borosilicate vapor cell contains a dispersive component as well. Upper bounds on the imaginary parts modeled as zero, as well as potential errors in general, are discussed below.
\begin{figure*}
\includegraphics[width=\linewidth, height=10cm]{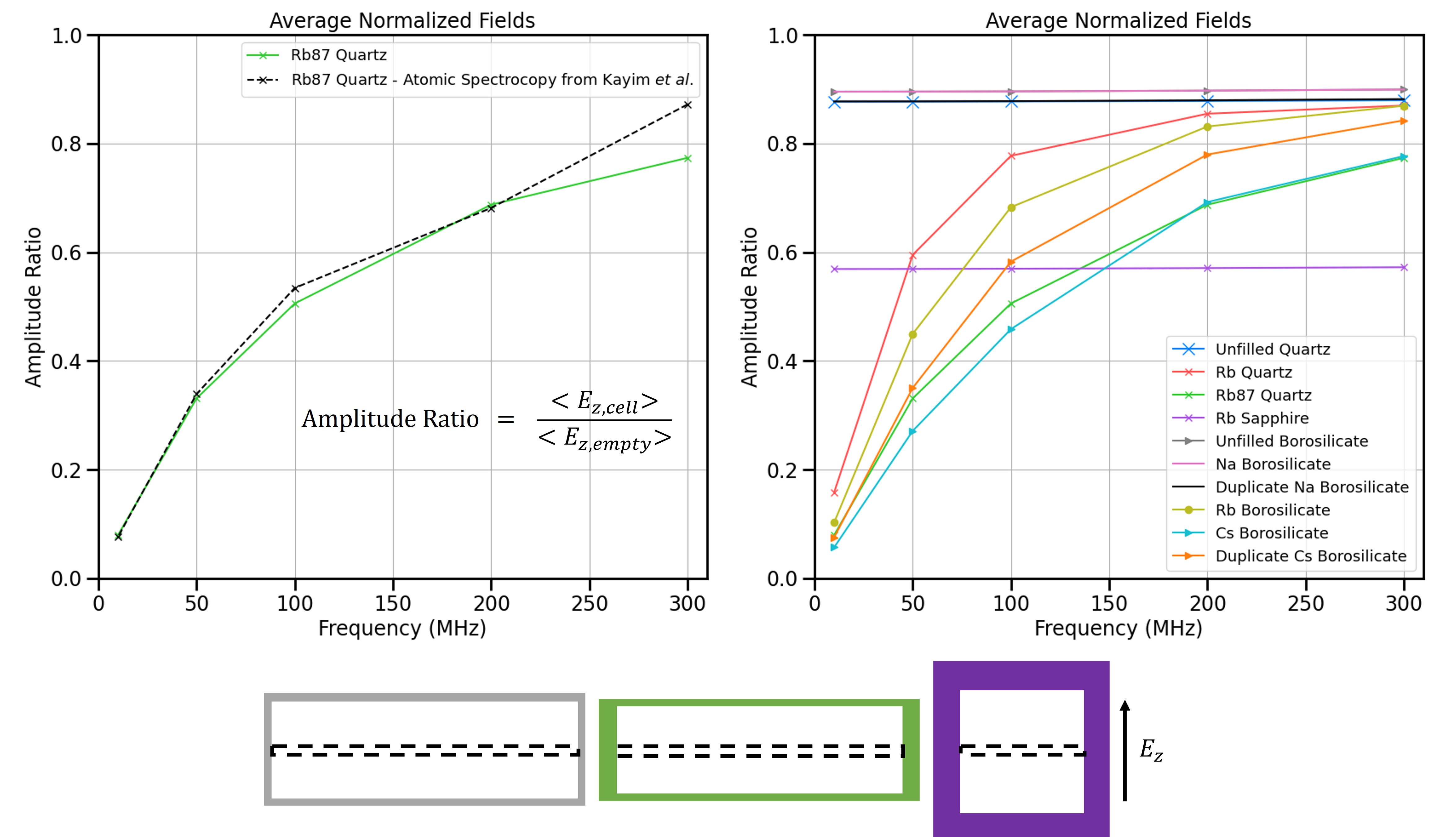}
\caption{Field reduction within the vapor cells due to the RF electromagnetic shielding effect. The electric field component normal to the waveguide plates is displayed next to each vapor cell from Fig.~\ref{fig_cells}. The fields are averaged across the highlighted hypothetical interior optical volume and normalized to the case of the same volume within the empty waveguide. The hypothetical optical beam width is approximately 1~mm in diameter.}
\label{fig_shielding}
\end{figure*}

\subsection{\label{Explanation Of Physical Mechanisms}Explanation Of Physical Mechanisms}
 In addition to quantifying the effective complex RF material properties, several additional observations can be made by comparing vapor cells. First, the dispersion and absorption observed is very likely due to the atomic vapor and vapor cell wall interaction. This is most evident here by comparing two cases: the rubidium-filled quartz vapor cells with the unfilled quartz vapor cell, and the rubidium-filled quartz vapor cells with the rubidium-filled sapphire vapor cell. The former suggests the response is not due to the quartz glass alone, and the latter suggests it is not due to the atomic vapor alone. This is further validated by comparing with the borosilicate vapor cells as well. Because the conductivity dominates the response in most of the filled vapor cells, this implies the majority of the response in all cells is due to the vapor and vapor cell wall interaction allowing free charges to readily flow between atoms adhering to the wall. However, the small dispersive response in two of the vapor cells also suggests a small contribution from a bound dipolar interaction within the atoms as well. This was anticipated and is mostly consistent with previous lower frequency measurements from ref.~\cite{sandia_new,sandia_old}, however these previous works did not observe or consider a dispersive component. Additionally, the sodium-filled borosilicate vapor cells showed little to no response, even though sodium itself is very conductive, indicating further that conductivity alone cannot explain this phenomenon. It should be noted that the sheet resistances determined from the conductivity values here (considering 1.5~mm wall thicknesses of quartz and borosilicate) are of similar magnitude to that of ref.~\cite{sandia_old}. Second, sapphire is likely the preferred material for the frequencies in this range. While it does have a naturally higher dielectric constant than quartz, it does not appear to experience significant dispersion or absorption, at least above 25~MHz. There is a more sizeable model and measurement mismatch for the sapphire vapor cell from 25 MHz, but this is likely a calibration or time gating induced error. The shielding factor of these vapor cells will be shape- and thickness-dependent, in addition to the effective complex RF material properties, so it is important to consider what is possible from a manufacturing standpoint as well. Next, though the two rubidium-filled quartz cells have identical vapor pressures (and similar for the cesium-filled and sodium-filled borosilicate vapor cells), they show a strong difference in effective complex RF material properties. One possibility is that the amount of vapor is different between the two vapor cells. It is also possible the distribution of vapor condensate on the walls within the vapor cell is different between vapor cells. These two possibilities will most certainly influence the outcomes. However, it should be noted that the authors examined the vapor cells visually, and with a microscope in some cases. There was no distinguishable difference in distributions or amounts of vapor on the walls, including sodium which had large amounts of vapor on the walls yet showed little response here. Similarly, the effect appears to be alkali dependent, with cesium showing a stronger response, however, it should be noted that there was variation between vapor cells for the same alkali. Any of these options suggest a possibility for controlling this effect. Some of the vapor cells are dispersive and other are purely absorptive. Because the two vapor cells showing dispersion had larger conductive responses as well, it is likely related to the amount of vapor or the interaction potential with the wall. Furthermore, it could be that the other filled vapor cells are in fact dispersive, but the effect is too small to capture here. If the effective complex permittivity model is extrapolated up to 1 GHz, i.e where $\epsilon_{\infty}$ dominates, it appears the effects are reduced to zero, though more work is needed to understand new potential effects at higher frequencies. As a final note, the sapphire vapor cell does have a borosilicate stem that in principle should experience an effect, even if the sapphire portion does not. When visually inspecting the cells, the sapphire vapor cell stem does not appear to contain a large amount of vapor and the stem itself contains a small portion of the volume. Thus, it is expected the stem places a minor role in the overall response, however this will likely become more important at lower frequencies when the interaction becomes stronger.

\subsection{\label{Error Analysis}Error Analysis}
It is difficult to report an accurate estimate for the potential error (i.e., there are no error bars) in the effective complex RF permittivity due to its complicated frequency-dependent relationship with the S-parameters. That is, it is difficult to accurately determine the true error in effective complex RF permittivity for a given S-parameter model and measurement mismatch for a given frequency and geometry. Additionally, there are potential measurement (calibration or time gating) errors or geometry modeling (Yee cell limitations) errors as well. To better understand the potential errors from each of these items, several studies were performed. First, several reference objects made of metal, alumina, Rexolite, and PTFE with varying curvature (round and sharp edges) were modeled and measured with very good agreement ($>$90$\%$ accuracy across the broadband). Additionally, as stated above, there was good agreement here when considering bulk dielectric properties of quartz, borosilicate, and sapphire. This leads the authors to believe the potential measurement or geometric modeling errors are qualitatively small. The S-parameter mismatch error can be quantified in several ways, but the most straightforward is to examine variations in fitting parameters. This error analysis was performed on the rubidium87-filled quartz glass vapor cell and the unfilled quartz glass vapor cell. By examining a 10$\%$ change in each of the fitting parameters, a bounds can be placed on the potential fitting error in regards to the S-parameter mismatch. The range of S-parameter values produced using these bounds is shown using a shaded region in Fig.~\ref{fig_variations}. In summary, the model and measurement consistency with a wide range of material, shape, and size, along with the error analysis for two of the vapor cells, leads the authors to believe it is reasonable to assume that the error of the extracted real and imaginary parts of the effective complex RF permittivity are within 10 percent. The conductivity of the rubidium-filled sapphire vapor cell is estimated to be less than 0.005 $S/m$, based on no noticeable changes in S-parameters below these values. This upper bound would indicate that if the rubidium-filled sapphire vapor cell does in fact show a conductive response, as is expected at lower frequencies, it would be orders of magnitude below that of the other conductive cases. Similarly, the conductivity of the borosilicate and quartz vapor cells with no loss term included in the model are estimated to be less than 0.001 $S/m$. For the unfilled vapor cells, this is consistent with bulk values for the dielectric loss of the various materials. 
\begin{figure}
\includegraphics[width=\linewidth]{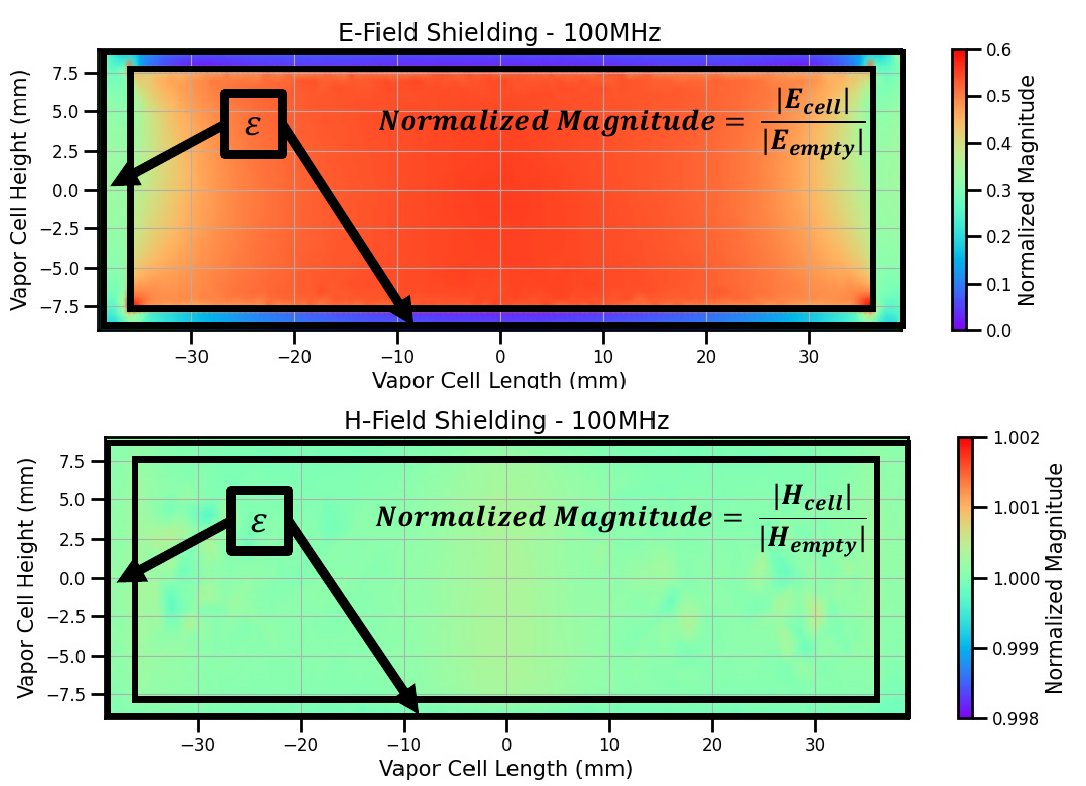}
\caption{Normalized magnitude of the electric and magnetic fields in a cross section centered on the rubidium87-filled quartz vapor cell. The magnitude of the fields is normalized to the case of the same area within the empty waveguide. This illustrates the field reduction and spatial degradation of the fields due to the effective complex RF material properties.}
\label{fig_shielding_2}
\end{figure}

\subsection{\label{RF Shielding Quantified}RF Shielding Quantified}
The RF electromagnetic shielding can be quantified several ways. To directly compare with ref.~\cite{Baran_paper}, and so to also provide further validation of the accuracy of the extracted properties, the electric field component normal to the waveguide plates, along the direction shown in Fig.~\ref{fig_shielding}, is averaged across the highlighted hypothetical optical volume. It is then normalized to the case of the empty waveguide using the same volume, and the optical beam is circular with a diameter of approximately 1~mm. As seen in Fig.~\ref{fig_shielding}, and described in more detail by Kayim \emph{et al.}, there is good agreement between the atomic-spectroscopy-determined electric field and the electric field determined using the extracted effective complex RF material properties for the rubidium87-filled quartz vapor cell, adding further validity and usefulness to the effective properties determined using this technique. In ref.~\cite{Baran_paper}, the authors demonstrate how this technique can be directly used to determined the intrinsic sensitivity of the atomic physics via field calibration, as well as the effective reduced sensitivity of the entire device due to the shielding of the packaging. 

All of the other vapor cells are also shown in  Fig.~\ref{fig_shielding} for a better understanding as to how the shielding changes with material, shape, and size. For example, though the rubidium-filled sapphire vapor cell shows no vapor to vapor cell wall interaction in this frequency range, the naturally larger dielectric constant with thicker walls still causes significant shielding. Similarly, though the unfilled borosilicate vapor cell has a higher dielectric constant than the unfilled quartz vapor cell, the overall shape and size, as well as the hypothetical optical beam distance from the walls, yields a smaller shielding effect. Because the sapphire vapor cell didn't show noticeable vapor to vapor cell wall interactions in this frequency range, it can be concluded that sapphire is a more suitable vapor cell material below around 50~MHz, assuming a different vapor wouldn't have led to a different result. The data also indicates above 50~MHz that quartz is a more suitable vapor cell material to reduce the impact of the shielding effect. Because sodium showed little to no impact across the broadband, a Rydberg sensor utilizing sodium would be a good choice in reducing this effect. This assumes in all cases that the atomic physics sensitivity is not effected by choice of alkali or vapor cell material. The S-parameters in Fig.~\ref{fig_S_param_compare} were measured from 15-300~MHz and modeled from approx. 15-285~MHz (based on the FDTD input pulse), while the data in Fig.~\ref{fig_Ep_real_and_imag} and Fig.~\ref{fig_shielding} show extracted properties from 10-300~MHz. Thus, the property extraction has been extrapolated a small amount for convenience of comparison. Additionally, the total magnitude of the electric field and magnetic field at 100~MHz within a centered cross section of the rubidium87-filled quartz vapor cell is shown in Fig.~\ref{fig_shielding_2}. Similar to above, the fields are normalized to the case of the empty waveguide (i.e. no vapor cell). The magnetic field is only slightly perturbed ($<$1$\%$), as expected. These field cuts illustrate, not only the field strength reduction, but also the spatial uniformity change as well, both of which effect sensitivity and direction finding.
\begin{figure}
\includegraphics[width=\linewidth]{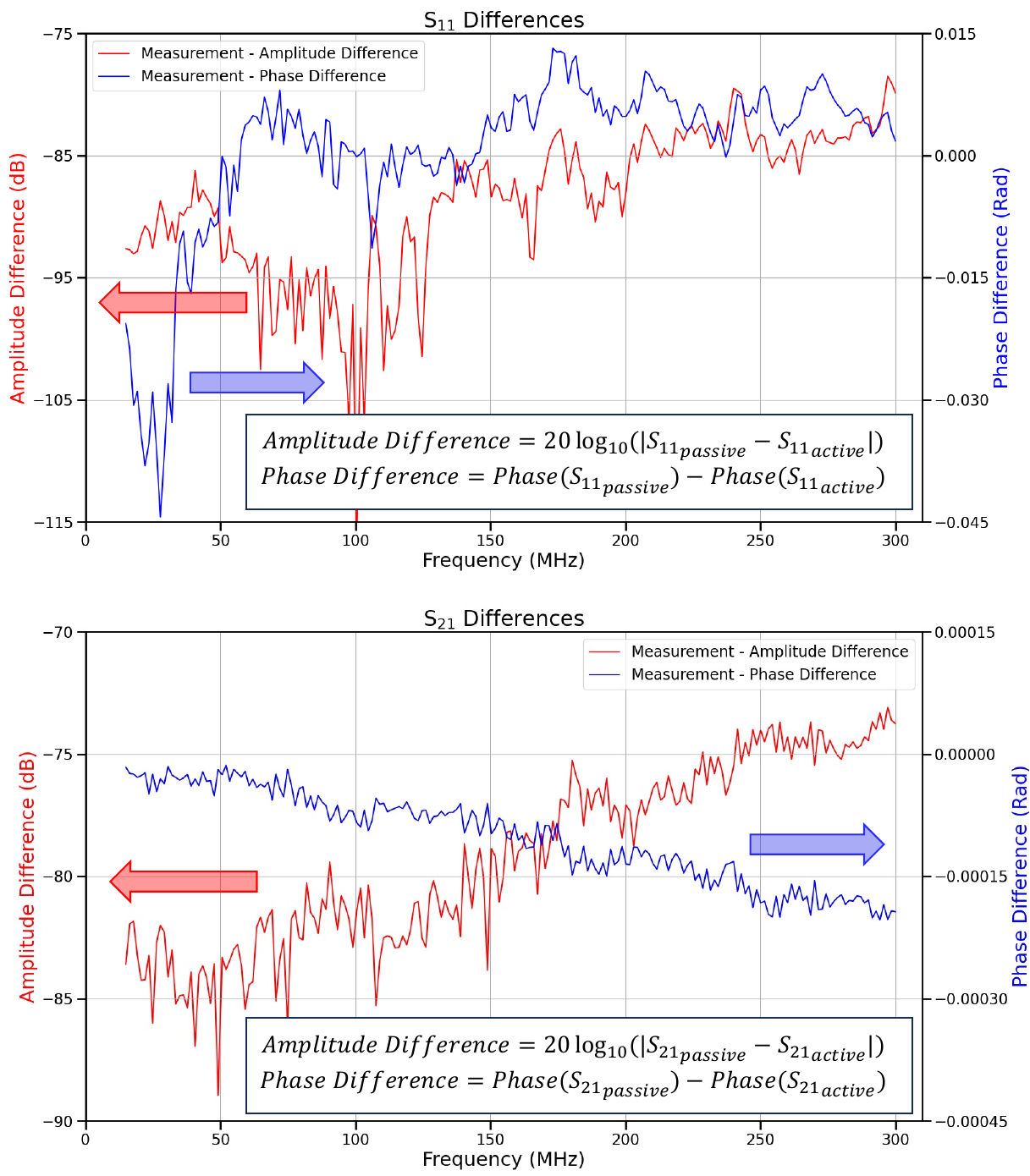}
\caption{Difference in $S_{11}$ and $S_{21}$ between the passive case of the rubidium87-filled quartz cell in Fig.~\ref{fig_S_param_compare} and the case where some of the vapor is operating in the active Rydberg state configuration of Ref.~\cite{Baran_paper}.}
\label{fig_difference}
\end{figure}

\subsection{\label{Impact of Act of Measurement on Permittivity}Act of Measurement Impact on Permittivity}
Lastly, a measurement was performed in which some of the Rydberg atoms were within the excited Rydberg state. Rydberg atom excitation-related contribution to this effect is also of great interest. However, at least in the configuration and vapor cell used in Kayim \emph{et al.} there was a very small, but likely measurable, change in S-parameters when the atoms were in the excited Rydberg state. The difference in the amplitude and phase of $S_{11}$ and $S_{21}$ is shown in Fig.~\ref{fig_difference} as a demonstration. This translates to a less than $1\%$ change in the magnitudes of $S_{11}$ and $S_{21}$ on average across the broadband. The effect on the effective complex RF permittivity can't be quantified here, but it can be concluded that the impact is very small in this case. Furthermore, this further validates the comparison in Fig.~\ref{fig_shielding}, since the act of measurement doesn't appear to substantially change the effective complex RF permittivity, and thus the shielding. It's important to note that only a small fraction of the total vapor is in the excited Rydberg state. Future studies that examine this effect in greater detail are of great interest, as it is likely this result will change with different atomic configurations or the frequency range considered.

\section{Conclusions\protect\\}
The atomic vapor and glass material wall interaction shielding of electromagnetic waves is well known, but not thoroughly studied across a wide frequency range, nor are the physical mechanisms completely understood. This effect makes it challenging for Rydberg sensors to accurately measure electric fields or to be used in arrays for direction finding, among other applications. This paper utilized a new method to quantify the effective complex RF material properties of several COTS vapor cells and their shielding amounts within the context of a stripline waveguide from 10-300~MHz. It was found the physical mechanisms can be understood as both a large electrically conductive response and a small Debye-like relaxation in some cases. That is, the vapor on the vapor cell wall allows both free charges to flow between the atoms on the wall and bound dipolar interactions. The effect is likely dependent on the amount of alkali, type of alkali, and placement of the alkali within the vapor cell. Applications of this technique include making precise numerical field corrections or physically designing a more optimal vapor cell via coatings, material changes, or geometric changes. 

\begin{acknowledgments}
The authors would like to thank Dr. John Burke and Dr. Jonathan Hoffman of the DARPA Microsystems Technology Office (MTO), as well as GTRI's Quantum Systems Division for materials, consultation, and atomic spectroscopy data. This research was developed with funding from the Defense Advanced Research Projects Agency (DARPA). The views, opinions and/or findings expressed are those of the author and should not be interpreted as representing the official views or policies of the Department of Defense or the U.S. Government. Distribution Statement A – Approved for Public Release, Distribution Unlimited.
\end{acknowledgments}

\bibliography{ref.bib}

\end{document}